\documentclass[12pt, a4paper]{article}

\usepackage{amsfonts}
\usepackage{amssymb}
\usepackage{amsmath}
\usepackage{array}
\usepackage{bm}
\usepackage{cancel}
\usepackage{cite}
\usepackage{epsfig}
\usepackage{graphicx, rotating}
\usepackage{hyperref}
\usepackage{latexsym}
\usepackage{multirow}
\usepackage{slashed}
\usepackage{verbatim}
\usepackage{xcolor}
\usepackage[export]{adjustbox}
\usepackage[normalem]{ulem}
\usepackage[utf8]{inputenc}

\newcommand{\be}{\begin{equation}}
\newcommand{\ee}{\end{equation}}
\newcommand{\bea}{\begin{eqnarray}}
\newcommand{\eea}{\end{eqnarray}}

\newcommand{\xidw}{\xi_{\rm DW}}

\begin{document}

\begin{titlepage}

\begin{flushright}
\small
DESY-23-020
\\
IFT-UAM/CSIC-23-14
\\
RESCEU-2/23
\\
TUM-HEP-1450/23
\end{flushright}
\vspace{.3in}

\begin{center}
{\Large\bf Gravitational waves from defect--driven}
\\[2mm]
{\Large\bf 
phase transitions: domain walls
}
\vskip 2mm
\bigskip\color{black}
\vspace{1cm}{
{\large
Simone Blasi$^a$,
Ryusuke Jinno$^{b,c}$,
Thomas Konstandin$^d$,\\[5pt]
Henrique Rubira$^e$, 
Isak Stomberg$^d$ 
}}

{\small
\vskip 5mm
$^a$ Theoretische Natuurkunde and IIHE/ELEM, Vrije Universiteit Brussel, \& The  International Solvay Institutes, Pleinlaan 2, B-1050 Brussels, Belgium\\
\vskip 1mm
$^b$ Instituto de F\'{\i}sica Te\'orica IFT-UAM/CSIC\\ 
C/ Nicol\'as Cabrera 13-15, Campus de Cantoblanco, 28049, Madrid, Spain
\vskip 1mm
$^c$ Research Center for the Early Universe (RESCEU), Graduate School of Science\\
The University of Tokyo, Tokyo 113-0033, Japan
\vskip 1mm
$^d$ Deutsches Elektronen-Synchrotron DESY, Notkestr.~85, 22607 Hamburg, Germany
\vskip 1mm
$^e$ Physik Department T31, Technische Universit\"at M\"unchen \\
James-Franck-Stra\ss e 1, D-85748 Garching, Germany\\
}

\bigskip

\begin{abstract}
We discuss the gravitational wave spectrum produced by first-order phase transitions seeded by domain wall networks. This setup is important for many two-step phase transitions as seen for example in the singlet extension of the standard model. Whenever the correlation length of the domain wall network is larger than the typical bubble size, this setup leads to a gravitational wave signal that is shifted to lower frequencies and with an enhanced amplitude compared to homogeneous phase transitions without domain walls. We discuss our results in light of the recent PTA hints for gravitational waves.

\end{abstract}

\end{center}

\end{titlepage}

\newpage
\section{Introduction}
\label{sec:Intro}

\vskip 1 cm

Cosmological first-order phase transitions (FOPTs) in the early universe can source a stochastic gravitational wave background (SGWB) which current and near-future observational searches could detect. The remarkable successes of LIGO-Virgo~\cite{LIGOScientific:2016aoc, LIGOScientific:2016sjg, LIGOScientific:2017vwq} to measure GWs from compact binaries in the kHz frequency band marks the beginning of an era of gravitational wave (GW) probes of the universe. While LIGO-Virgo and other interferometer-based GW searches continue to be developed around the world, the pulsar timing arrays and the associated experiments, including NANOGrav, EPTA, PPTA, and IPTA, are simultaneously making strides towards detecting a stochastic background and have recently reported evidence for a stochastic common-spectrum process among the pulsars~\cite{NANOGrav:2020bcs, Chen:2021rqp, Goncharov:2021oub, Antoniadis:2022pcn}. While further evidence is necessary to conclusively confirm or reject the SGWB hypothesis, it has nevertheless been shown that current observations could possibly be explained by GWs from a FOPT~\cite{NANOGrav:2021flc, Xue:2021gyq}. 

Next in line we find LISA, the Laser Interferometer Space Antenna, designed to measure GWs in the mHz frequency band~\cite{eLISA:2013xep, Babak:2017tow, LISACosmologyWorkingGroup:2022jok}. While the strongest candidate to source GWs in this band are supermassive black-hole binaries, LISA offers to probe GWs from FOPTs occurring around the electroweak scale (TeV). The possibility to measure the GW spectrum over various frequency bands calls for accurate predictions of GW spectra that could discriminate various cosmological scenarios. 

In this paper, we investigate the influence of defects on the dynamics of FOPTs and their 
GW production (see~\cite{Steinhardt:1981ec,
Steinhardt:1981mm,Jensen:1982jv,Hosotani:1982ii,
PhysRevD.30.272} for early works on impurities,~\cite{Hiscock:1987hn,Green:2006nv,Gregory:2013hja,Burda:2015yfa,Mukaida:2017bgd,Canko:2017ebb,Dai:2019eei,El-Menoufi:2020ron,Oshita:2018ptr,Balkin:2021zfd,Dai:2021boq} for the case of black holes and local over--densities, and 
\cite{Preskill:1992ck,
Yajnik:1986tg,Yajnik:1986wq,Dasgupta:1997kn,
Kumar:2008jb,Lee:2013zca,Lee:2013ega,Koga:2019mee,
Kumar:2009pr,Kumar:2010mv,Agrawal:2022hnf,
Dunsky:2021tih, Strumia:2022jil} for strings, monopoles, and defects in general).
One interesting possibility is that a FOPT responsible for the stochastic GW production is locally catalyzed, or \emph{seeded}, by a domain wall (DW) network ~\cite{Blasi:2022woz}. Depending on the underlying particle physics model, the nucleation 
probability can be significantly larger on the surfaces of the DWs than in the bulk of the Universe. 

The presence of a DW network effectively introduces a new length scale, $\xidw$, denoting the mean separation of the DWs. Whenever $ \xidw$ exceeds the mean separation of adjacent bubbles, $\langle r \rangle$,  (see Appendix \ref{app:separation} for details) 
\be
\xidw \gg 2 \langle r \rangle \simeq 6 v_w/\beta \, ,
\label{eq:limit}
\ee
we expect that the network imprints a signature in the GW spectrum.
In particular, we expect that the peak frequency is shifted to lower frequencies and that the peak is enhanced, akin to GW spectra computed from FOPTs modified by macroscopic thermal fluctuations as discussed in Ref.~\cite{Jinno:2021ury}. We furthermore expect that the spectral shape is modified due to the spatially inhomogeneous bubble distribution as induced by the DW network; rather than colliding bubbles, the DW network results in the propagation of sheets moving away from the walls which collide after the mean time $\xidw/v_{w}$. 

Notice that in the extreme limit (\ref{eq:limit}), the bubble nucleations can be regarded as essentially simultaneous everywhere on the DW network compared to the timescale for bubbles to travel between the walls, and $\beta$ becomes an irrelevant parameter. 
Moreover, the dynamics of the DWs can be neglected in this limit. 
Since the resulting GW spectrum will still be well-defined in this limit,
all relevant features in the nucleation history and the GW spectrum cannot
depend on $\beta$ for large DW networks. This point will be discussed in detail in Section~\ref{sec:Ising}. 
By virtue of the arguments above, we only need a snapshot of the DW network with an appropriate 
correlation length to infer a possible \emph{bubble nucleation history}, i.e. the bubble nucleation sites and times, and the corresponding GW spectrum. Instead of proper simulations of DW networks, we will mimic them with a Monte Carlo implementation of the Ising model. 
Using Monte Carlo Methods to study DWs was already put forward in the classic 
review on DWs~\cite{Vilenkin:1984ib} and more generally in the analysis
of percolation~\cite{Stauffer:1978kr}.
The above procedure should at least qualitatively yield a reasonable distribution 
of bubble nucleation sites. This topic will be discussed further in Section~\ref{sec:Ising}. 

Once a bubble nucleation history is obtained, we use hydrodynamic simulations to compute the GW signal following~\cite{Jinno:2022mie}. The naive expectation is, again, that once the DW network becomes relevant in the limit (\ref{eq:limit}), the GW spectrum peak frequency is reduced and the amplitude enhanced compared to a conventional FOPT without a DW network. The last point fits well with the observation that the GW production is proportional to the correlation length of the energy-momentum tensor. In this study, we aim to explore these dependencies and how the overall shape of GW spectra is affected by a DW network. We also discuss our results in light of the recent PTA hints for gravitational waves.

\vskip 0.4 cm

The outline of the manuscript is as follows:
In Section \ref{sec:Ising} we discuss the relevant properties of the DW networks. We also 
provide some results for the Ising model (in the low-temperature regime) and how to make 
contact with DW networks using the correlation length of the system. 
In Section \ref{sec:GWs} we discuss the gravitational wave spectra produced by our hydrodynamic 
simulations and discuss the results in Section~\ref{sec:dis}.

\section{Domain walls}
\label{sec:Ising}

In this section, we start by describing the DW formation mechanism and discuss the regimes in which DWs can have relevant phenomenological consequences. Later, we argue why the Ising model resembles a realistic setup to mimic configurations of DW networks from which bubble nucleation histories of a FOPT seeded by such a network can be deduced.

\subsection{Formation and dynamics of domain walls}

DWs arise whenever a discrete symmetry is broken during the course of the Universe~\cite{Kibble:1980mv}.
The dynamics of such DWs has been studied numerically since several 
decades (see for example~\cite{Press:1989yh, Garagounis:2002kt, Hiramatsu:2010yz, Kawasaki:2011vv, Hiramatsu:2013qaa}).
The consensus is that sometime after the FOPT, the DW
network enters a scaling regime where the correlation length of the system 
approaches the Hubble parameter and the energy density roughly scales as $\rho \sim H \sim 1/t$~\footnote{There is 
an argument in the literature whether there are logarithmic corrections to this scaling, see~\cite{Hindmarsh:1996xv, Garagounis:2002kt}.}.
In the long term, this is disastrous, since the DW network would begin 
to dominate the energy density. Moreover, the DWs would induce fluctuations in the 
plasma. Both these effects are in conflict with CMB observations.
In order to alleviate these issues, the DWs have to disappear before they 
reach a sizable fraction of the 
energy density of the Universe. 
Several solutions are conceivable: 
\begin{itemize}
\item i) The discrete symmetry is explicitly slightly broken~\cite{Sikivie:1982qv}. This leads to different vacua that are not entirely degenerate and 
the global minimum will dominate after a while.
\item ii) There is a bias in the initial conditions~\cite{Larsson:1996sp}. Also in this scenario, depending on the initial conditions, the DWs will decay
 before they become harmful.
\item iii) The exact discrete symmetry is initially spontaneously broken. Later, the symmetry is again restored whereby the DWs cease to exist. 
\end{itemize}
In all these scenarios, the DWs will produce gravitational waves~\cite{Gleiser:1998na, Hiramatsu:2010yz, Kawasaki:2011vv, Hiramatsu:2013qaa, Saikawa:2017hiv}. 
The setup we have in mind is of type iii), more concretely we envision a two-step phase transition
where the second step is of first order. 
Consider the potential 
\be
V = \frac\lambda4 (\phi^2 - v^2)^2 + \frac{\lambda_m}4 s^2 \phi^2 + \frac{\lambda_m}4 s^4 
- \frac12 \mu_S^2 s^2 \, ,
\ee
where $\phi$ is a Higgs field and $s$ a gauge singlet endowed with a $Z_2$ symmetry. For some 
regions in parameter space, this system will have a two-step phase transition: first the 
system transitions to a phase where $s$ obtains a VEV. This breaks the $Z_2$ symmetry 
spontaneously and produces DWs. At somewhat lower temperatures, the system
attains the ground state of the system with a finite $\phi$-VEV but unbroken 
$Z_2$ symmetry ($\left< s \right>=0$). The DWs cease to exist once the system 
completely transitions into the low-temperature phase. This avoids the over-closure issue and 
excessive fluctuations in the plasma~\cite{Espinosa:2011ax, Patel:2012pi}. 

In this setup, the DWs exist only for a relatively short period of time.
Ultimately, the gravitational waves are  not predominantly produced by the DWs or the 
scalar field that drives the phase transition, but by sound waves that are generated during the phase 
transition. To quantify the spectrum of GWs, we use hydrodynamic 
simulations. Compared to conventional FOPTs, the nucleation history of bubbles is modified since bubbles preferentially nucleate on the DWs~\cite{Blasi:2022woz}. This turns out to be a generic feature in all the two-step parameter space, where the temperature for the seeded FOPT is always higher than the would-be homogeneous nucleation temperature~\cite{Blasi:2022woz}. Intuitively, a faster nucleation rate around the DWs can be understood by noticing that the energy stored in the DW tension can be used to tunnel through the potential barrier. 

A crucial parameter controlling the phenomenology of the FOPT is the number of DWs per Hubble patch at the time of seeded nucleation,
or equivalently the correlation length of the system which describes the average curvature radius as well as the mean distance among the walls. In the picture we are considering, DWs are formed during the first step of the FOPT when the $Z_2$ symmetry is spontaneously broken.
If this transition is of the second-order type, the typical size of domains at formation will be given by the microscopic scale associated with the singlet mass $\xi_\text{DW} \sim 1/m_s$. 

The subsequent evolution depends strongly on whether the DW motion is damped by friction or if instead the walls can freely oscillate at (mildly) relativistic velocities. In the latter case the DW network is expected to increase its correlation length to saturate the causality bound from the Hubble expansion, yielding $\xi_\text{DW} \sim t$\,\cite{Press:1989yh}. This corresponds to the standard scaling regime where one expects $\mathcal{O}(1)$ DWs per Hubble volume.
On the other hand, when friction dominates the correlation length of the system, it is supposed to increase at a slower pace. Standard arguments based on a balance between the tension force, $\sigma_\text{DW}/\xi_\text{DW}$, and the friction force that scales with
$T^4 v$ (where $v$ is the DW velocity), yielding to $\xi_\text{DW} \sim (G \sigma_\text{DW})^\frac{1}{2} t^\frac{3}{2}$, where $G$ is the Newton's constant \cite{Vilenkin:2000jqa}. Following this estimate, the correlation length of the system will still be $\xi_\text{DW} \ll H^{-1}$ at the time of seeded nucleation, as the two steps of the phase transition are usually expected to be separated at most by a factor of a few in temperature~\cite{Blasi:2022woz}. In this case, the phenomenology of the FOPT will resemble the standard homogeneous case in which the duration is controlled by the time dependence of the nucleation rate.

In our analysis we will not engage in a model--dependent determination of the number of DWs at the seeded nucleation temperature, but rather use the correlation length of the network as an input to determine the phenomenological implications for the GW spectrum. In doing so we will consider an intermediate case in which the DW size has grown to macroscopic scales but the standard scaling regime has not yet been reached,
so that $\xi_\text{DW} < H^{-1}$. As the correlation length of the network always 
stays smaller than the Hubble size, the expansion of the Universe can be neglected in the hydrodynamic simulations of the GW production. Ultimately, the expansion of the Universe might only lead to a moderate 
suppression of the GW signal, but neglecting the expansion simplifies the numerical implementation of the simulation.

\vskip 0.4 cm

From simulations of DWs, it is known that the main properties of
DW networks are encoded in a single quantity: the correlation length. 
Instead of simulating a full DW network, it is tempting to mimic 
it with a simpler system with similar properties. The prototypical example of this 
is the Ising model. Close to the critical temperature, the system displays universality and 
is applicable to many systems with phase transitions. 
Here we rely on the fact that the outcome of the Ising model
resembles very well how DWs appear in genuine simulations of DW networks.  
There are several arguments why a snapshot of the Ising model should suffice for our purpose.
Whether a single snapshot of the DW network provides enough information 
to simulate the bubble nucleation history, or if alternatively the dynamics of the DW network is relevant. 
Simulations in the scaling regime show that DWs can oscillate with relativistic velocity.
From the first nucleation of bubbles until the end of the FOPT, the bubble wall of the new phase has 
to travel the correlation length of the DW network, $\Delta t \sim \xidw/v_w$. However, the time window
during which actual bubble nucleations happen is rather small, $\Delta t \sim 10/\beta$.
Hence, as long as one considers the limit (\ref{eq:limit}), the motion of the DWs 
is irrelevant.

In principle, there could be an issue with DWs
propagating faster than the velocity of the bubble wall interfaces, $v_w$, changing the bubble nucleation history in the process. However, in the broken (new) phase, the DWs cease to exist, since 
the $Z_2$ symmetry is restored. Overall, a snapshot of the DW network should suffice to deduce a realistic bubble nucleation history. 

\subsection{Modeling the domain wall network}\label{sec:modeling}

Even if the typical correlation length of the DW network coincides with
 the one from the Ising model, one might wonder to what extent the substructure of the two systems agree. For example, small islands could be present in the DW network whose size distribution is not captured by the scale of the large walls. These are, for instance, generated when two big walls collide and a closed wall is produced in the aftermath. These walls would further collapse, producing small structures (`spheres' or `bags'), similar to loops in networks of cosmological strings. 

A large number of closed islands would have a strong impact on the gravitational wave production, as in this case the duration of the FOPT would be controlled by the average distance between the islands rather than the distance between the "infinite" walls.
However, those substructures are not observed in sizable numbers in simulations (see ~\cite{Press:1989yh} and \cite{Garagounis:2002kt} for visual representations). In fact, differently to the case of cosmic strings, the energy losses allowing the DW network to reach the scaling regime are most likely due to particle radiation (see also Ref.\,\cite{Martins:2016ois}  where it was found that the `chopping' parameter for a DW network is indeed compatible with zero).

In order to generate the actual bubble nucleation history from the Ising model, one has to map this Ising model data to a 
realistic DW network. As mentioned before, the details of this procedure, as we will see, are not essential, since in the limit (\ref{eq:limit}) (when we expect the GW signal to be 
affected), the bubble nucleation happens simultaneously everywhere on the network. This is because the time scale of the 
FOPT is much shorter than the separation of the DWs in this limit, in which case the 
broken phase will grow uniformly out of the DW network with a velocity 
given by the usual propagation speed of the bubble walls $v_w$. In practice,
 the correlation length of the DWs will only be one order larger than the mean bubble size
 and we will explore two different possibilities for this mapping (see Appendix~\ref{app:isingparam}).

\begin{figure}
	\centering
	\includegraphics[width=0.24\textwidth]{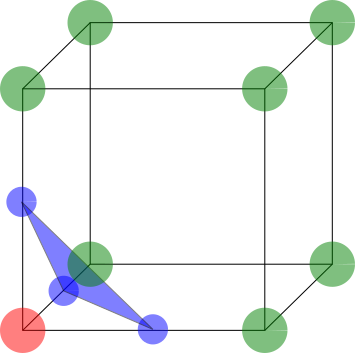} 
	\includegraphics[width=0.24\textwidth]{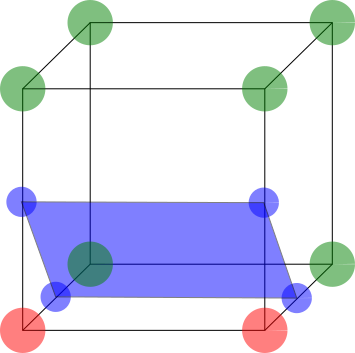} 
	\includegraphics[width=0.24\textwidth]{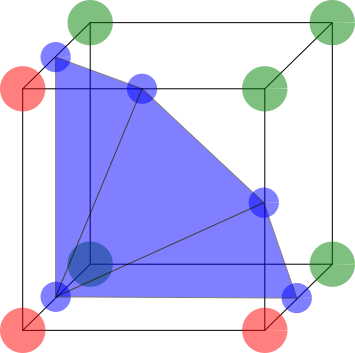} 
	\includegraphics[width=0.24\textwidth]{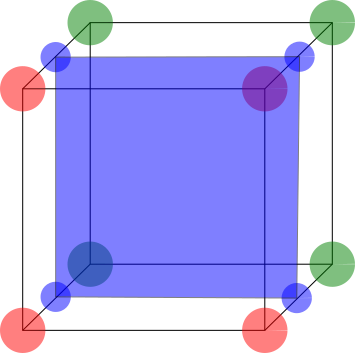} 
	\caption{Some representative DW configurations for cells with 0 to 4 corners in the opposite phases. The 
		corresponding surface weights of the DW network are $\{ 0, \sqrt{3}/8, 1/\sqrt{2}, 1/\sqrt{2} + \sqrt{11}/8,1\}$. The representative of the empty cell is omitted.
	}
	\label{fig:surfaces}
\end{figure}

Ideally, one would like to assign to every grid cell in the simulation a
{\em weight} that represents the surface area of the DW network in the cell. We implemented two different methods to assign weights to the grid densities, both leading to identical results. We now introduce one of those methods. The presentation and comparison with the second (and to some extent simpler) method is delivered in Appendix~\ref{app:weigth}.

One, and perhaps the most straightforward, way to assign weights to cells is by observing for every cell the state of the 
eight corners and then constructing a surface that separates the two phases. 
Such a surface runs through the midpoints of the edges between corners of opposite states (see Fig. \ref{fig:surfaces}).

While, in principle, every corner configuration is possible, only those which are maximally connected are likely to occur, the others being suppressed by virtue of the Ising model. A slight simplification to the method of constructing surfaces through midpoints is, therefore, to assume that corners of equal states are most compactly connected. 
We hence chose a configuration that is representative of all cells of the same corner counts.   
Identification of the weight can then proceed by means of simply counting the number of corners in each phase since in this case, the mapping between corner counts and weight is one-to-one. Consider $k,l$ corners in each phase. Then, the corresponding weights $w_{kl}$ are $w_{08} = 0$, $w_{17} = \sqrt{3}/8$, $w_{26} = 1/\sqrt{2}$, $w_{35} = 1/\sqrt{2} + \sqrt{11}/8$, and $w_{44} = 1$, which are proportional to the enclosed DW area. While this simplification induces a systematic error in the estimation of the DW area, we show in Appendix \ref{app:weigth} that the final results are insensitive to this error. The probability to nucleate a bubble in cell $ijk$ at time $t$ is then proportional to $w_{ijk}\exp(t \beta)$, where $w_{ijk}$ is the weight of cell $ijk$, whereby a bubble nucleation history can be constructed as in the standard picture without DWs (see Sec 2.1 of \cite{Jinno:2022mie}). Such bubble nucleation histories will serve as inputs to the hydrodynamic simulations, which will be discussed in the next section. Notice that, since $w_{ijk}\neq 0$ only for cells containing a DW surface, bubbles nucleate along the DWs only.

\begin{figure}
	\centering
	\includegraphics[width=0.4\textwidth]{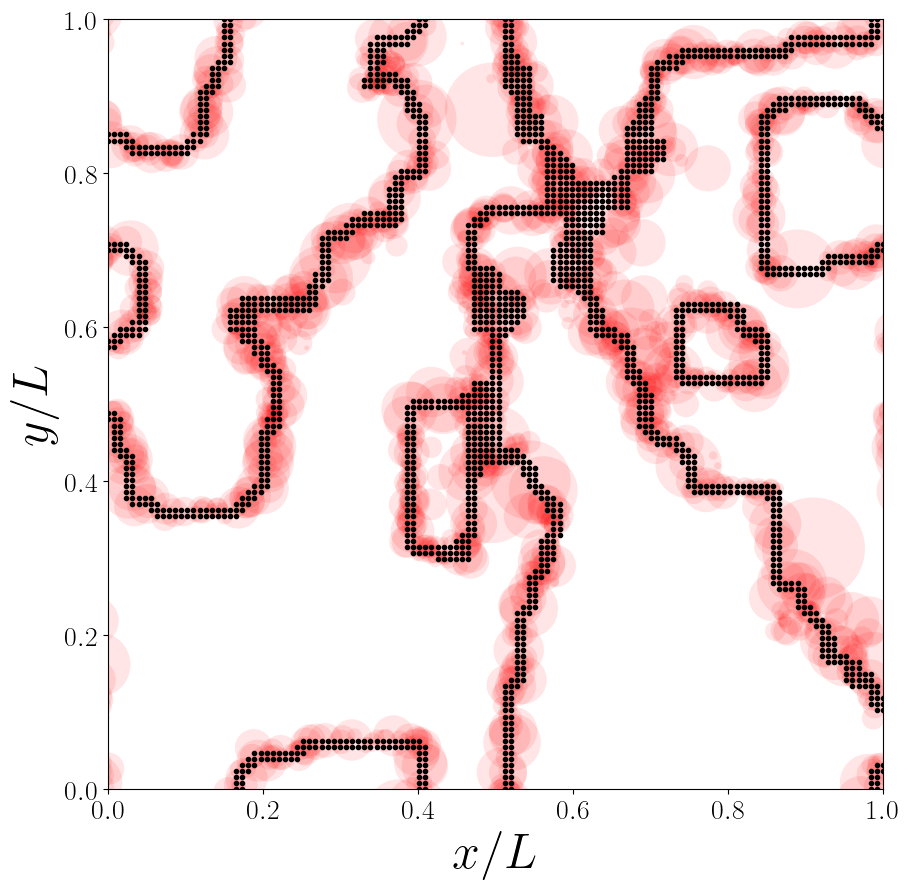} 
	\includegraphics[width=0.43\textwidth]{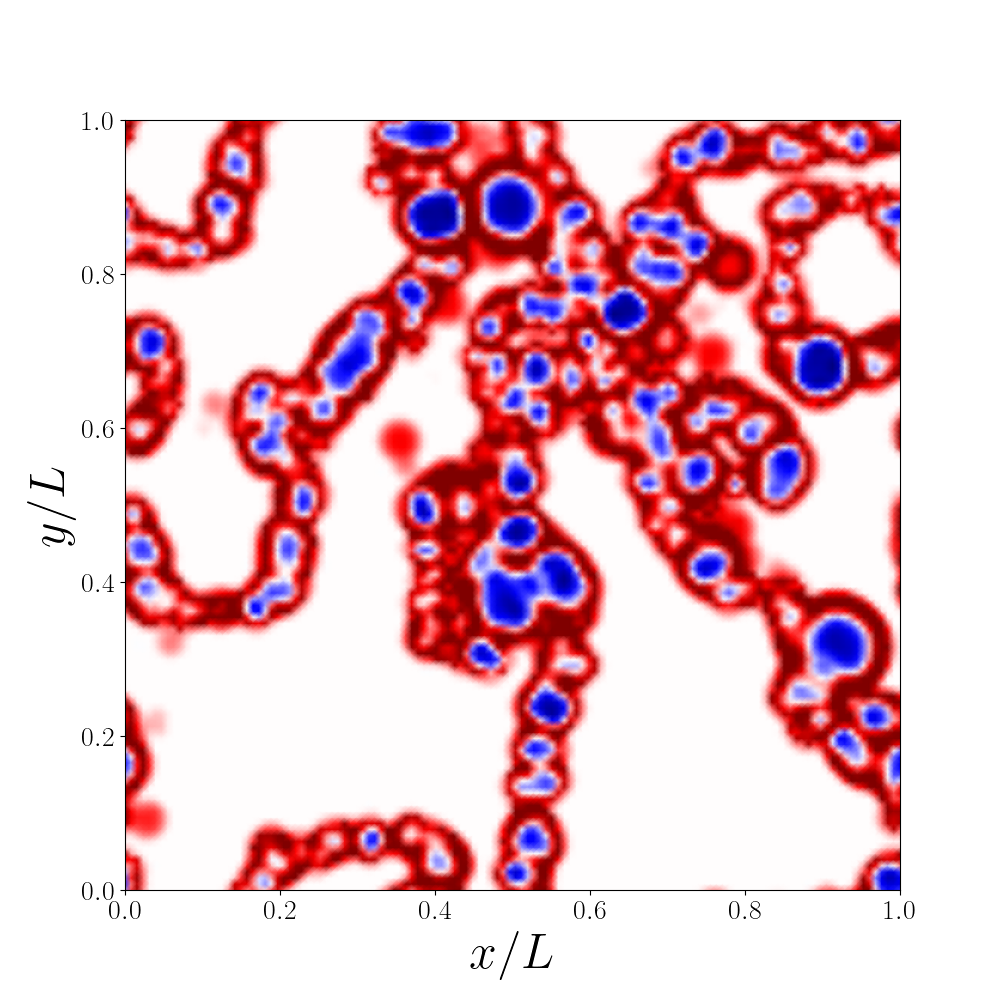} \\
	\includegraphics[width=0.4\textwidth]{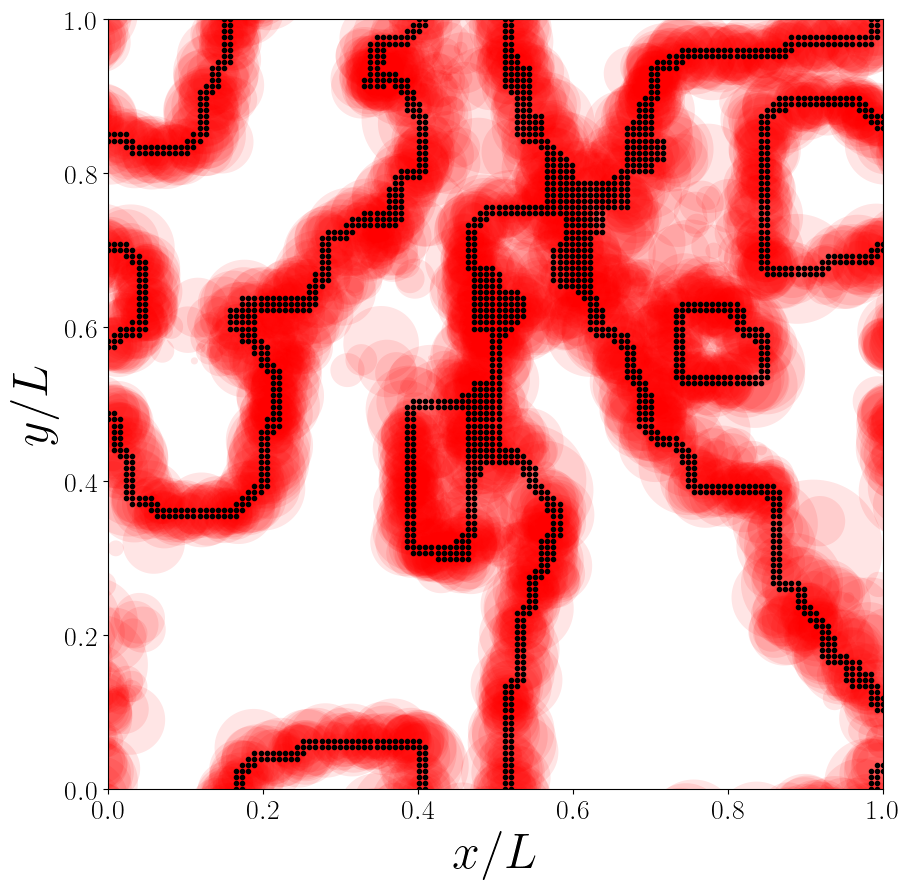} 	
	\includegraphics[width=0.43\textwidth]{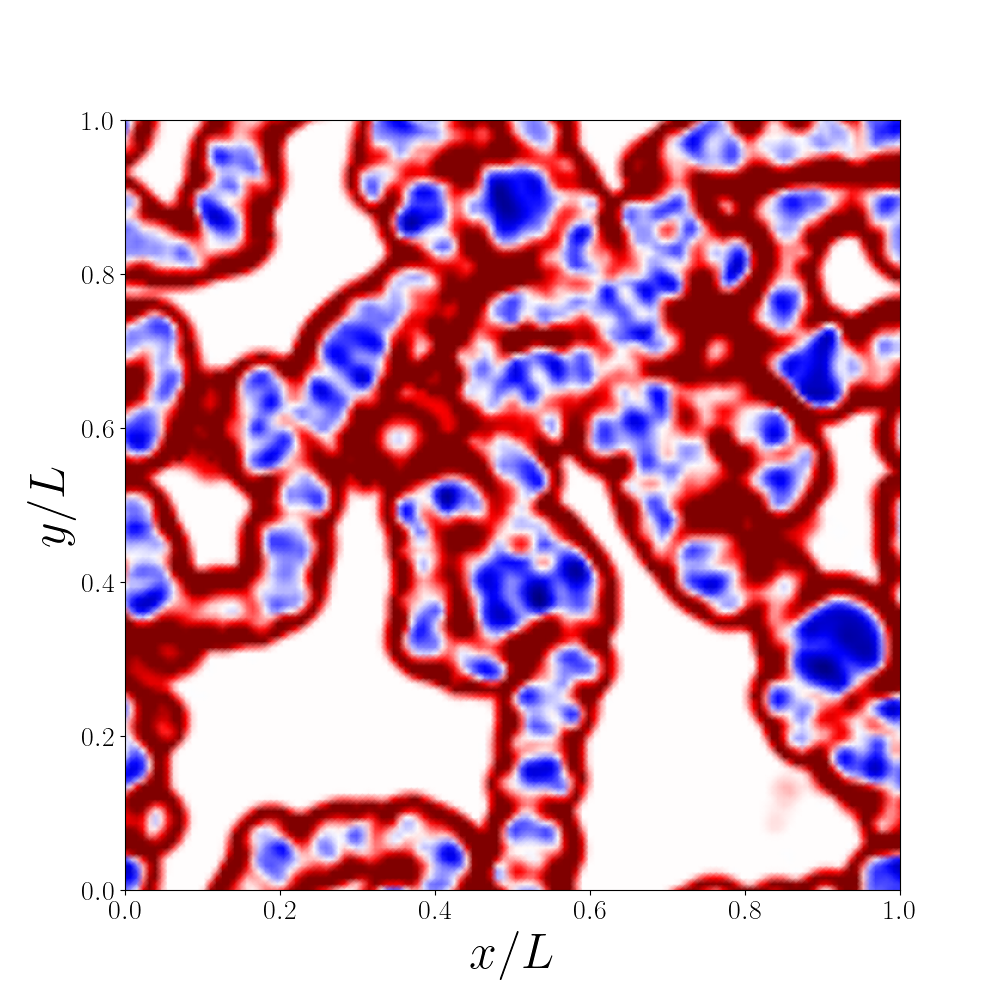} \\
	\includegraphics[width=0.4\textwidth]{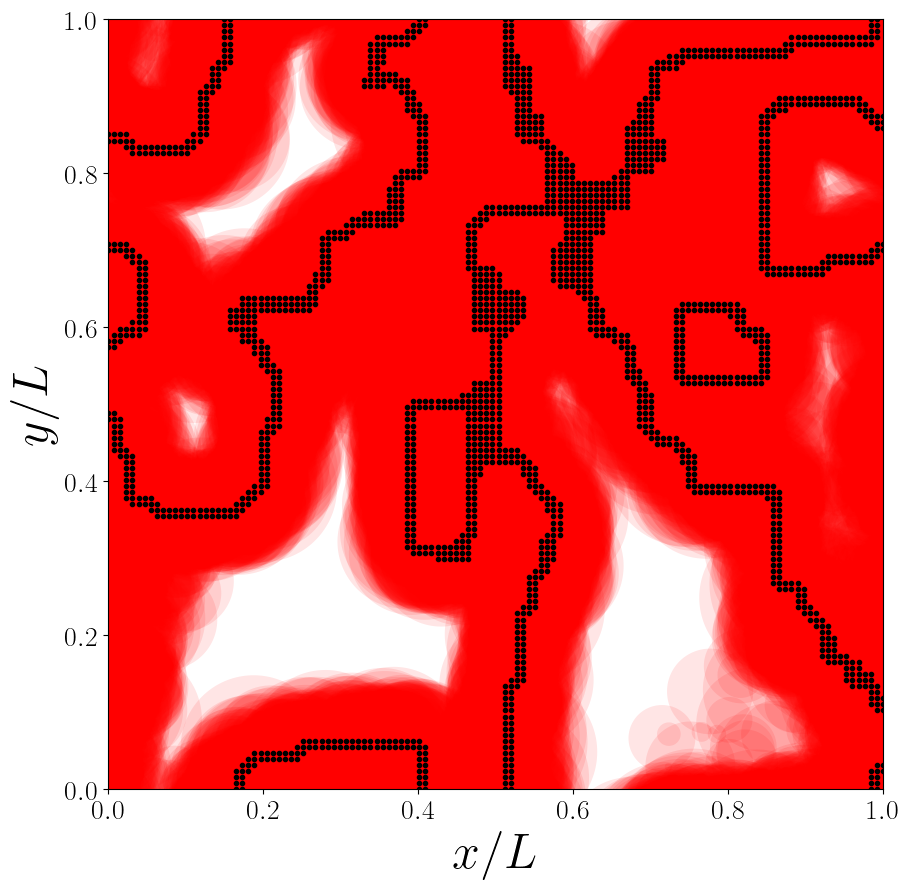} 	
	\includegraphics[width=0.43\textwidth]{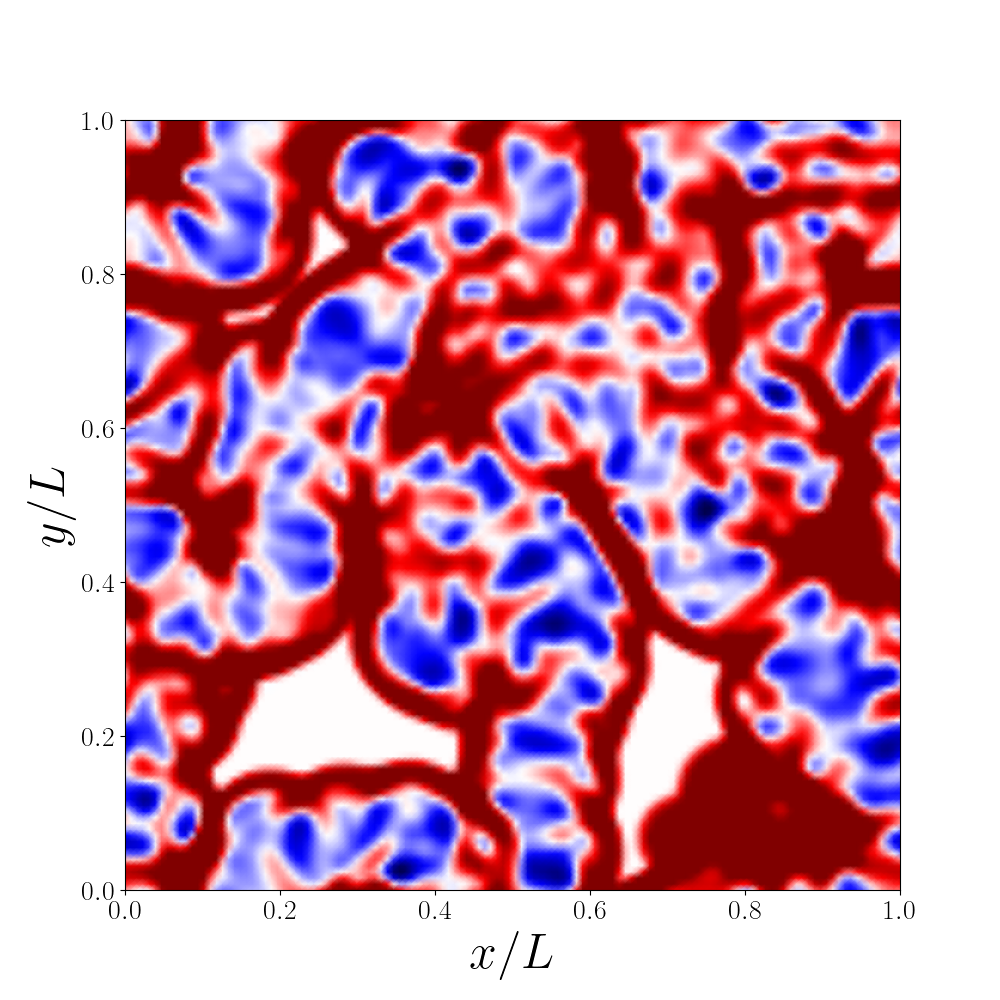}	 
	
	\caption{The figure shows a slice of the simulation of the Ising model at various time steps. The left panel shows the interface between the two domains and nucleated bubbles while the right panel displays the kinetic energy in the fluid. The used parameters are $v_w=0.8$, $L = 160 v_w/\beta$, $\xidw = 0.1 L$.}
	\label{fig:DWsnap}
\end{figure}

As emphasized before, the DW network is characterized by the mean separation between the DWs, $\xidw$.
One way to estimate this measure is to consider the total surface of the DW network, $S_{\rm DW}$.
The mean separation is then approximately
\be
\xidw \sim \frac{V}{S_{\rm DW}}
\label{eq:si_surf}
\ee
where $V = L^3$ denotes the volume of the simulation. For example, for the DW network in Fig.~\ref{fig:DWsnap},
we obtain $\xidw \sim 0.1 \times L$, where $L$ denotes the box size of the simulation. Starting from a random initial state and using the Metropolis algorithm, we perform $4\times 10^8$ phase alterations on a grid of size $128^3$ with the parameter $J/T=2$ to yield this correlation length (see App.~\ref{app:variants}).

In this estimate, the total surface area of the 
DW network was extracted using the weights discussed above. Alternatively, one can extract the correlation length using the correlation function
\be
\xi(\vec x) = \left< \phi(\vec y) \phi(\vec y + \vec x)\right> \, .
\ee
The field $\phi$ denotes the order parameter of the DW network that attains $\pm v$ in the two
degenerate minima. Moreover, one can study correlations in Fourier space leading to the power 
spectrum of the order parameter
\be
P_\phi(\vec k) = \int d^3x \, \xi(\vec x) \, \exp(-i  \, \vec x \cdot \vec k )\,  .
\ee

For large volumes the system should be rotationally symmetric such that $P_\phi(k)$ and $\xi(x)$ only depend 
on absolute values and are related by 
\be
\xi(x) = \frac{1}{2\pi^2}\int dk\, k^2 P_\phi(k) \frac{\sin(k x)}{kx}\, .
\label{eq:xiP}
\ee
In order to build intuition and make the connection with the correlation length $\xidw$, we discuss some 
limiting cases of these quantities. The contributions to the correlation function $\xi(x)$ are $\pm v^2$, depending on whether the order parameter $\phi$ is in the same vacuum at $\vec y$ and $\vec x + \vec y$ or not.
Hence, as long as the thickness of the DWs is neglected, 
one obtains $\xi(0) = v^2$.  

For increasing $x$ the correlation function decreases. As long as $x \ll \xidw$, the curvature of the DW
can be neglected and the contribution to the correlation function depends on the probability that the locations 
$\vec y$ and $\vec x + \vec y$ are on different sides of the DW. A simple estimate gives 
\be
\xi(x) \simeq v^2 \left(1 - \frac{|x| S_{\rm DW}}{V} \right) = 
v^2 \left(1 - \frac{|x|}{\xidw} \right) \, ,
\label{eq:xi_estimate}
\ee
in the limit $x \ll \xidw$, while in the large scale limit $x \gg \xidw$, $\xi(x) = 0 $. 
On the other hand, the power spectrum $P_{\phi}(k)$ is expected to go to a constant for $k\to 0$ 
and 
vanish for $k\to \infty$ (because the DW domains have no substructures). Notice that the relation (\ref{eq:xiP})
seems at first sight at odds with the fact the $\xi(x)$ has a linear term in (\ref{eq:xi_estimate}). The correlation function in (\ref{eq:xiP}) is even, $\xi(-x) = \xi(x)$, and expanding the sine function seems to imply that the correlation function can only have even powers of $x$. There is however a caveat in case the 
power spectrum does not falls off faster than $k^{-4}$ in the UV. In this case the correlation function is still even, $\xi(-x) = \xi(x)$, but the expansion of the $\sin$ function is not valid. 
In particular, the correlation function will develop a kink at $x=0$.

In order to extract the correlation length, we use for the power spectrum 
\be
P(k) \propto \frac{k_0^4}{k_0^4 + \kappa k^2 + k^4} \, .
\ee
In the case $\kappa = 2k_0$, one finds $\xi(x) = v^2 \exp(- x/\xidw)$,
while in the general case, this yields a correlation function of the form
\be
\xi(x) = v^2 \exp(- x/\xidw) \frac{\sin(x/x_o)}{x/x_o}  \, .
\label{eq:xi_fit}
\ee

\begin{figure}
	\centering
	\includegraphics[width=0.496\textwidth]{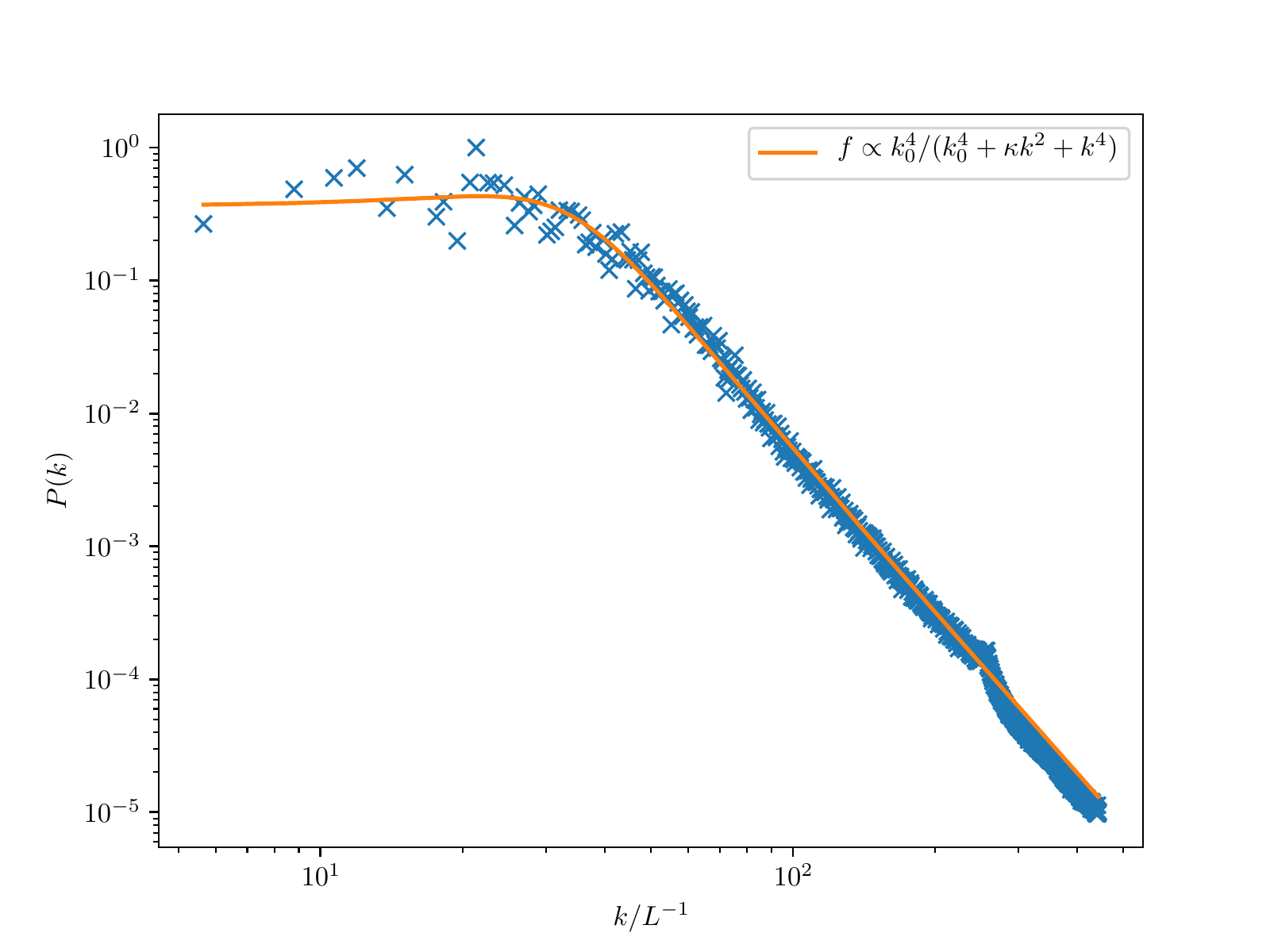} 
	\includegraphics[width=0.496\textwidth]{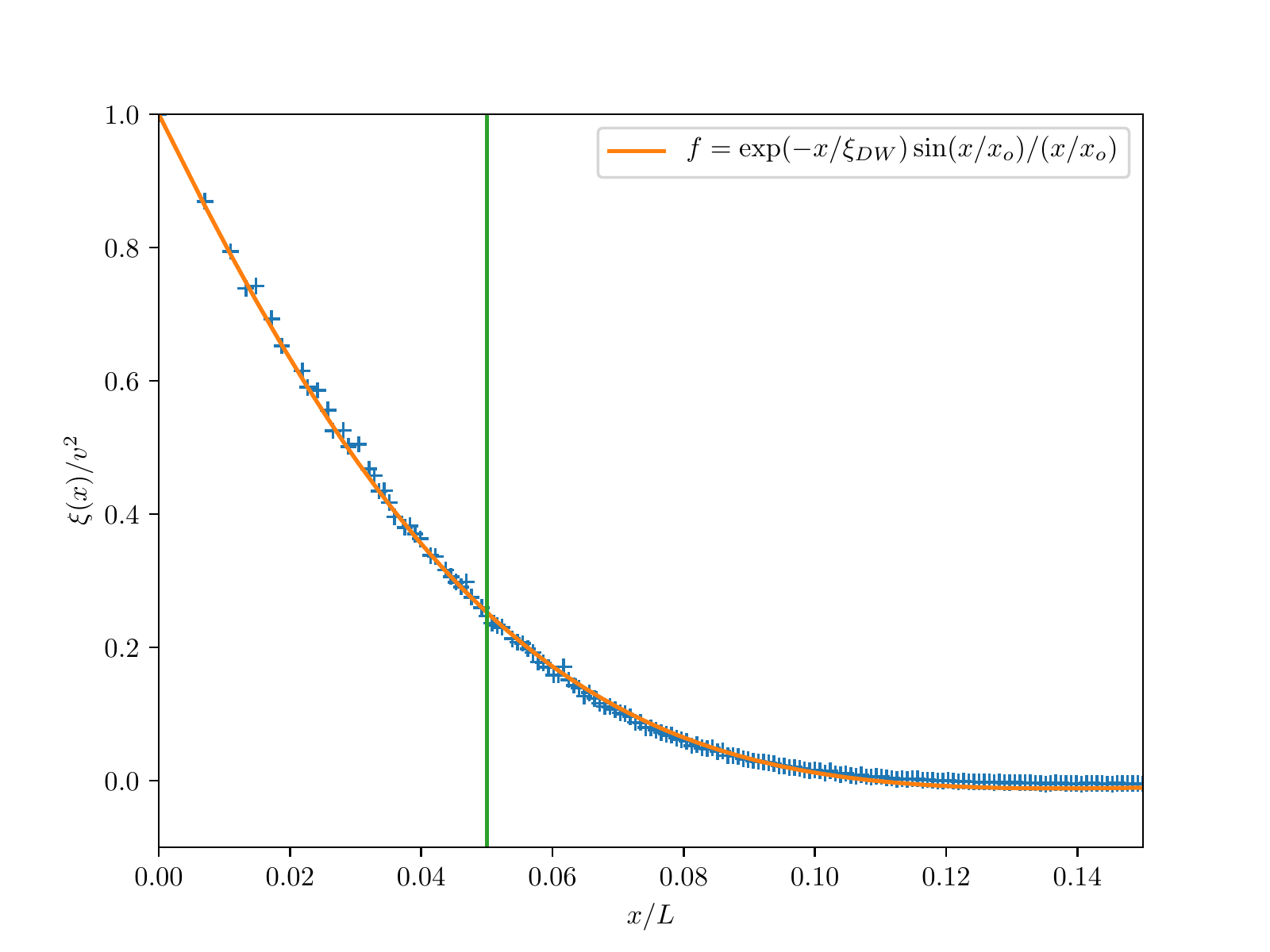} 
	\caption{The power spectrum, $P(k)$, and correlation function $\xi(x)$ of the Ising model in arbitrary units. The  fit can be used to extract the correlation length, $\xidw = 0.0499 L$, very close to the value inferred from the surface of the DW network, $\xidw = 0.05 L$.  
	}
	\label{fig:power}
\end{figure}

In Fig.~\ref{fig:power} we show the power spectrum and correlation function of the DW network. 
We also show the fits according to the shapes we expect. The extracted correlation length for the fit in (\ref{eq:xi_fit}) is $\xidw = 0.0499  L$ compared to the value $\xidw = 0.05  L$ that 
is obtained via relation (\ref{eq:si_surf}). The scale related to the oscillations turns out to be $x_o=0.035L < \xidw$ which leads to a slightly negative parameter $\kappa$ in the power spectrum. Notice that we used a smaller correlation length in these 
simulations to allow for more domains and better statistics in our measurement. In the end, the power spectrum analysis shows that the correlation length of the Ising model encodes all relevant information and resembles the DW system.

\section{Simulations of first-order phase transitions}
\label{sec:GWs}

To perform hydrodynamic simulations of FOPTs, we use the \emph{Higgsless} approach, as recently presented in~\cite{Jinno:2022mie}. In this approach, the scalar field is not dynamical but acts as a background that changes the equation of state as a bubble wall passes over a certain location during the FOPT. The fluid responds to this change in the equation of state whereby latent energy is transferred from the vacuum to the fluid and self-similar bubble profiles develop. The bubble locations and times are predetermined by the bubble nucleation history according to an exponential-in-time nucleation probability and the bubble wall velocity is a fixed external parameter. The advantage of this approach is that the bubble wall does not need to be resolved in the simulation as long as the numerical algorithm remains stable and converges despite the presence of shocks in the plasma. Overall, this provides a very efficient framework to predict the GW spectra of FOPTs.

We calculate the GW spectrum in the simulations using the dimensionless GW power~\cite{Jinno:2020eqg}
\be
Q'(q) = \frac{q^3 \beta}{w^2 V T} \int \frac{d\Omega_k}{4 \pi} 
\left[ \Lambda_{ij,kl} T_{ij} (q,\vec k) T_{kl}^* (q,\vec k) \right]_{q=k} \, ,
\ee
with $w$ representing the enthalpy in the false vacuum. The frequency and momentum of the GW waves are represented respectively by $q$ and $\vec k$, and $T$ is
the simulation time. The tensor $\Lambda$ projects the energy-momentum tensor $T_{ij}$ on its transverse traceless part. The observed GW power spectrum is then given by  
\be
\frac{\Omega_{\rm GW}}{Q'} = \frac{4H \tau_{\rm sw}}{3\pi^2} \frac{H}{\beta} \, ,
\label{eq:OmQp}
\ee
which is directly proportional to the lifetime of the sound waves $\tau_{\rm sw}$ (determined either by the onset of turbulence or Hubble damping).

In the presence of DWs, we use relatively large box sizes, $L = 160 v_w/\beta$,
to make sure that we are in the regime where the mean separation between the DWs is 
larger than the mean bubble size in the FOPT without the DWs~\footnote{
In the simulations, it is most efficient to scale the box size in the different runs with
 the wall velocity. This way, the number of nucleated bubbles (which is important to limit boundary effects) 
is kept constant and even bubble nucleation histories can be reused across simulations with 
different wall velocities~\cite{Jinno:2020eqg, Jinno:2022mie}. Likewise, we scale the correlation length $\xidw$ with the box size and hence also the wall velocity.
}. 

\begin{figure}
        \centering
        \includegraphics[width=0.45\textwidth]{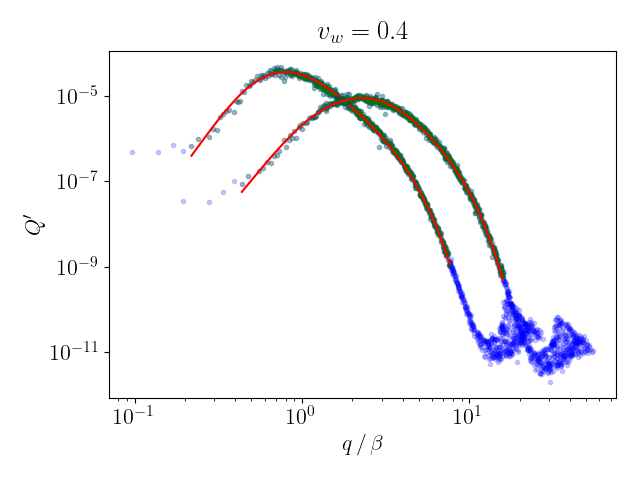} 
        \includegraphics[width=0.45\textwidth]{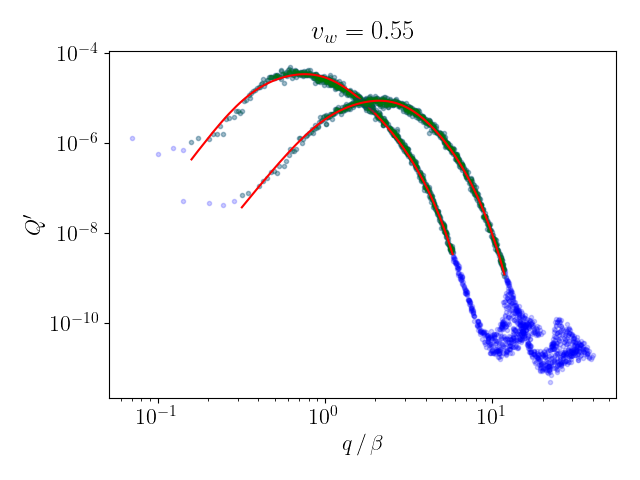} 
        \includegraphics[width=0.45\textwidth]{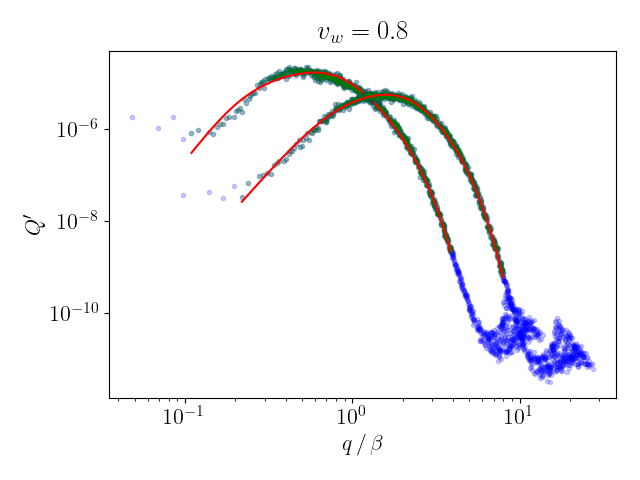} 
        \caption{Final spectra of the gravitational waves with (left) and without (right) a DW network. 
          The strength of the FOPT is $\alpha=0.05$, and the velocities of the 
bubble walls are $v_w = 0.4$, $0.55$ and $0.8$. The green points indicate the part of the spectrum 
that is used in the fit (shown in red). 
        }
        \label{fig:spectra}
\end{figure}

As mentioned before, the FOPT then happens quickly enough that the separation of the DWs will be imprinted in the GW spectrum. 
Figure \ref{fig:spectra} shows the final spectra of the simulations. 
The simulations with (without) bubbles have been run with box sizes $80 v_w/\beta$ ($160 v_w/\beta$) 
and simulation durations $64/\beta$ ($128/\beta$) in order 
to accommodate the proper size of the nucleated bubbles at percolation. 

As expected, the 
GW signal is shifted to smaller frequencies and enhanced by a similar factor. 
Moreover, the spectrum seems somewhat steeper in the IR. In order to extract the features of the spectra, 
we use a similar family of functions as in~\cite{Jinno:2022mie}. 
\be
S_f(q) = S_0 \times \frac{(q/q_0)^n}{1 + (q/q_0)^{(n-1)} [ 1 + (q/q_1)^4 ]} \times e^{-(q/q_e)^2} \, .
\label{eq:fit}
\ee
Compared to~\cite{Jinno:2022mie}, the Ansatz is slightly generalized to allow for an IR behavior 
different than $q^3$. 
Since the FOPT is relatively strong, there is no clear separation 
of the scales related to the bubble size and the shell thickness, so $q_0 \sim q_1$ in all fits
and $q_0$ is basically the peak of the spectrum.

Table \ref{tab:summary} summarizes our quantitative findings. $\Omega$ denotes the 
frequency-integrated GW power, while $q_0$ and $n$ are fitting parameters according to (\ref{eq:fit}).
Quantities with a bar are computed from simulations without DWs. 

\begin{table}
\centering
  \begin{tabular}{ l || c | c | c ||c ||c ||c |c }
    $v_w$ & $q_0/\beta$ & $\xidw / \beta^{-1}$ & $q_0 \cdot \xidw$  & $q_0/\bar q_0$ & $\Omega/ {\bar\Omega}$  & $n$& $\bar n$ \\
    \hline
    \hline
    0.4 & 0.50 & 6.4 & 3.19  & 3.69 & 2.37 & 5.51 & 4.87\\ 
    \hline
    0.55 & 0.35 & 8.8 & 3.07 & 3.95 & 2.75 & 5.03 & 4.63\\ 
    \hline
    0.8 & 0.22 & 12.8 & 2.81  & 3.40 & 3.44 & 4.78 & 4.14\\
  \end{tabular}
\caption{Parameters extracted from the spectra in Fig.~\ref{fig:spectra}.
Quantities with a bar are computed without DWs. The frequency $q_0$ is obtained fitting Eq.~(\ref{eq:fit}) to simulation data. The domain wall correlation length $\xidw$ is obtained from the Ising model snapshots (notice that it scales with $v_w$). 
}
  \label{tab:summary}
\end{table}
The peak of the spectrum closely correlates with the correlation length of the 
DW network, $q_0 \simeq 3/\xidw$. As expected, in presence of DWs, the peak is shifted to smaller frequencies while the 
amplitude is enhanced by a similar factor. In the fits, the IR tail is somewhat
steeper but all simulation boxes are in principle too small to reliably 
extract the slope of IR behavior $n$.

\section{Discussion and conclusions}
\label{sec:dis}

We studied the gravitational wave spectrum from a first-order phase transition seeded by domain walls.
When bubbles nucleate preferentially on the domain walls, the characteristic 
length scale will be the correlation length of the domain wall network if it is larger than the 
typical bubble size in a homogeneous first-order phase transition. In fact, the inverse duration of the 
first-order phase transition, $\beta$, will cease to be relevant once the correlation length of the domain walls network is   
large enough.

Accordingly, the correlation length of the sound waves in the plasma will be 
increased. This moves the peak frequency to smaller values and enhances the 
peak amplitude by a similar factor. Overall, a reasonable estimate for the gravitational wave spectrum is 
given by using the established formulae for the spectrum from a first-order phase transition and making the 
replacement 
\be
\beta \to 3/\xidw \simeq  4 \, {\rm max} (v_w,c_s)/\xidw \, ,
\ee
by just matching the mean bubble size of the homogeneous first-order phase transition to the correlation length of the domain walls (see Fig.~\ref{fig:spectra} and Table~\ref{tab:summary}). 
The factor on the right side takes into account that for detonations the bubble size relates to the 
wall velocity, while for deflagrations the shock always propagates with the speed of sound.
Moreover, the IR tail of the spectrum appears a bit steeper in our simulations than in conventional 
first-order phase transition. However, the IR tale of the spectrum suffers from quite large 
statistical uncertainties and the effect is still tentative right now. 
Also notice that the temperature of the seeded phase transition is somewhat higher since 
the seeded nucleation probability dominates over the homogeneous one~\cite{Blasi:2022woz}.

Finally, let us comment on a possible application of this effect. 
Recently, NanoGrav and other pulsar timing array experiments hinted at a signal of a stochastic gravitational 
background~\cite{NANOGrav:2020bcs, Chen:2021rqp, Goncharov:2021oub, Antoniadis:2022pcn}. One possible explanation for such a signal could be a 
first-order phase transition around a few MeV ~\cite{NANOGrav:2021flc, Xue:2021gyq}. However,
a cosmological first-order phase transition at such low temperatures tends to produce tension with 
the number of effective degrees of freedom measured in big bang nucleosynthesis~(see e.g.~\cite{Ramberg:2022irf}). 
In many models, the strength and duration of the first-order phase transition are correlated.
If the first-order phase transition is seeded by domain walls and the 
gravitational wave spectrum peaks at the correlation length of the domain wall network, 
this correlation can be broken and the first-order phase transition
temperature can be accordingly somewhat larger. 
This can allow explaining the pulsar timing array excess without being in tension with big bang nucleosynthesis constraints.

\section*{Acknowledgment}

HR acknowledges Oriol Pujol\`{a}s and Diego Blas for helpful discussions. 
SB is thankful to Prateek Agrawal, Alberto Mariotti and Michael Nee for useful discussions. SB is supported by FWO-Vlaanderen through grant number 12B2323N. HR is supported by the Excellence Cluster ORIGINS, which is funded by the Deutsche Forschungsgemeinschaft (DFG, German Research
Foundation) under Germany’s Excellence Strategy - EXC-2094 - 390783311. 
TK and IS are supported by the Deutsche Forschungsgemeinschaft (DFG, German Research Foundation) under Germany's Excellence Strategy -- EXC 2121 ``Quantum Universe" -- 390833306. 
We acknowledge the support of the European Consortium for Astroparticle Theory in the form of an Exchange Travel Grant.

\appendix

\section{Statistical quantities on the bubble distribution\label{app:separation}}

In this appendix, we estimate several statistical quantities related to the bubble distribution.

The mean bubble separation can be calculated from the expected number of bubbles $N_b$ nucleating in a box of volume $V$.
The differential probability $dP$ for a bubble to nucleate in the infinitesimal volume $d^3x$ around $\vec{x}$ in the time interval $[t, t + dt]$ is
\begin{align}
dP
&=
{\rm (prob.~for~}
\vec{x}
{\rm ~to~be~in~the~false~vacuum)}
\nonumber \\
&\quad \times
{\rm (prob.~for~one~bubble~to~nucleate~in~}
[t, t + dt]
{\rm )}
\nonumber \\
&=
\exp \left[ - \int_{- \infty}^t dt'~\frac{4\pi}{3} v_w^3 (t - t')^3 \Gamma(t') \right]
\times \Gamma(t) \,dt \, d^3x.
\end{align}
Using $\Gamma (t)=\Gamma_*e^{\beta (t - t_*)}$, one obtains 
\begin{align}
N_b
&=
\int dP
=
\frac{\beta^3 V}{8 \pi v_w^3}.
\end{align}
Thus the mean separation is
\begin{align}
R_*
&=
\left( \frac{V}{N_b} \right)^{\frac{1}{3}}
=
(8 \pi)^{\frac{1}{3}} \frac{v_w}{\beta}
\simeq
\frac{3 v_w}{\beta}.
\end{align}

The average bubble radius $\langle r \rangle$ can be calculated from the distribution of bubble radii when an arbitrary spatial point is passed by a wall for the first time.
To derive it, consider the differential probability $dP$ for an arbitrary spatial position $\vec{x}$ to be passed for the first time by a bubble with radius $r$ in the time interval $[t,t+dt]$.
It is written as
\begin{align}
dP
&=
{\rm (prob.~for~}
\vec{x}
{\rm ~to~be~in~the~false~vacuum)}
\nonumber \\
&\quad \times
{\rm (prob.~for~one~bubble~to~nucleate~in~}
[t - r, t - r + dt]
{\rm ~on~the~past~cone)}
\nonumber \\
&=
\exp \left[ - \int_{- \infty}^tdt'~\frac{4\pi}{3} v_w^3 (t - t')^3 \Gamma(t') \right]
\times
4\pi r^2 \Gamma(t - r / v_w) \, dt.
\end{align}
Again using $\Gamma (t) = \Gamma_* e^{\beta (t - t_*)}$, one obtains the radius distribution
\begin{align}
P (r)
&=
\int dP
=
\frac{\beta^3}{2v_w^3} r^2 e^{-\beta r / v_w}.
\end{align}
Thus the average bubble radius is, therefore, estimated as
\begin{align}
\langle r \rangle
&=
\int dr~r P(r)
=
\frac{3 v_w}{\beta}.
\end{align}

\section{Variants of the Ising model\label{app:variants}}

In this section, we study the effect of parameter choices for the Ising model and DW weight assignments to assess to what extent 
our choices in the main text influence our findings.

\subsection{Ising Model parameters}\label{app:isingparam}

\begin{figure}
        \centering
        \includegraphics[width=0.60\textwidth]{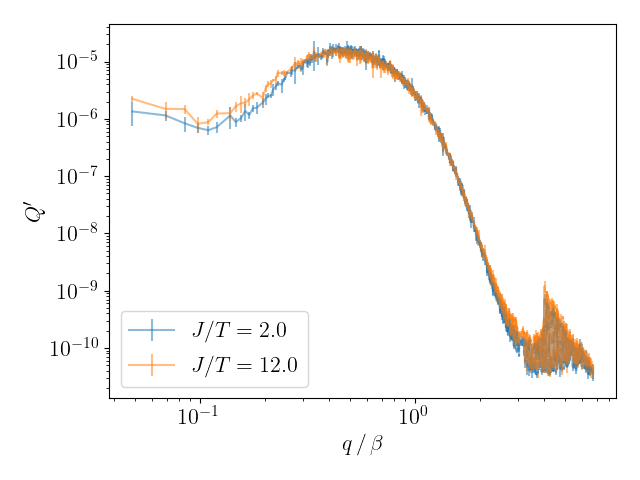} 
        \caption{Variants of the Ising model. The orange data corresponds to $J/T=12$ while the blue data uses $J/T=2$.
 		For each choice, we average over three different (randomized) initial conditions of the DW network. The error shows the variance between these three realizations with the same parameters.  
        }
        \label{fig:seedsJoT}
\end{figure}

In Fig.~\ref{fig:seedsJoT} we show the dependence of the GW spectrum on the value of $J/T$. In Fig.~\ref{fig:seedsJoT}, we show the dependence of the GW spectrum on the value of the Ising model parameters, the interaction $J$ and the temperature $T$ (or more concretely on the ratio $J/T$). The Hamiltonian of the system is given by 
\be
H = \sum_{i,j} J\, \sigma_i\, \sigma_j \, ,
\ee
where $\sigma_i = \pm 1$ denotes the state on site $i$.

The orange data corresponds to $J/T=12$ while the blue data uses $J/T=2$. For each choice, we average over three different (randomized) initial conditions of the DW network. The error shows the variance between these three realizations with the same parameters. 
The correlation length of the DW network was chosen to be relatively large, $\xidw = 16v_w/\beta$, and $v_w = 0.8$,
such that it dominates the bubble distribution according to (\ref{eq:limit}).

\subsection{Cell weights assignment}\label{app:weigth}

\begin{figure}
        \centering
        \includegraphics[width=0.496\textwidth]{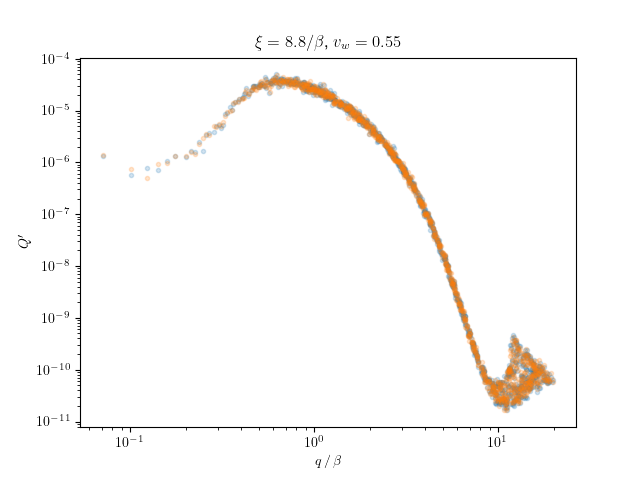} 
        \includegraphics[width=0.496\textwidth]{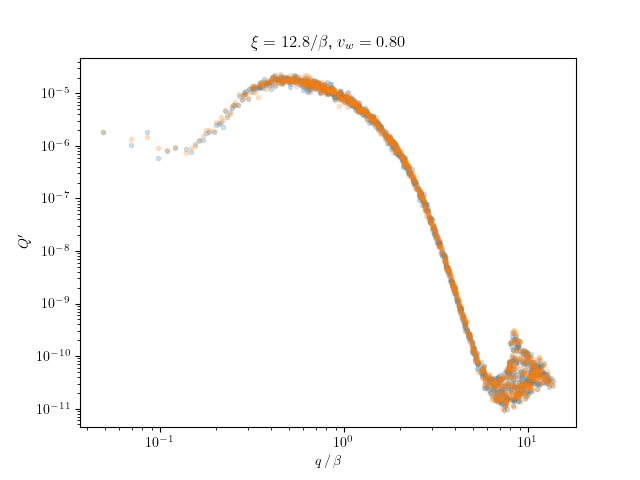} 
        \caption{Different methods to assign DW weights to the grid cells. The method from the main text is shown in orange, and the simplistic method described in \ref{app:weigth} in blue. 
        }
        \label{fig:TvH}
\end{figure}

Another possible modification of our methods concerns the mapping from Ising model to a DW network and corresponding cell weight assignment. 
In the main text (Sec.~\ref{sec:modeling}), we argued for a relatively simple procedure using just the 
count of the states for the eight corners in each cell. 
An even simpler model consists of just weighting the DW network proportionally to the number of corners of opposite phases in the individual cells of the Ising model simulation. For example, in the case that 
of the eight corners 0 to 4 corners are of opposite states, one could simply assign the values 
0 to 4 to the weights. Various GW spectra comparing the two different methods for the weights are shown in~\ref{fig:TvH}, for various parameter values.
Differences are always minute.

\newpage

\bibliographystyle{JHEP}
\bibliography{refs}

\end{document}